\documentclass[twocolumn]{aastex631}

\def\gsim{\ \raise 3pt \hbox{$\rangle$} \kern -8.5pt \raise -2pt \hbox{$\sim$}\ }
\usepackage{graphicx,natbib,color,amsmath,subfigure}
\usepackage{longtable}
\usepackage{ulem}
\newcommand{\blank}[1]{}
\usepackage{color}
\begin{document}

\newcommand{\pr}{^\prime}
\newcommand{\pc}{\tl{pc}}
\newcommand{\kpc}{\tl{kpc}}
\newcommand{\Mpc}{\tl{Mpc}}
\newcommand{\ha}{H$\alpha$}

\newcommand{\vect}[1]{\mbox{\bf #1}}

\newcommand{\dotp}{\!\cdot\!}
\newcommand{\crs}{\!\times\!}

\newcommand{\be}{\begin{equation}}
\newcommand{\ee}{\end{equation}}
\newcommand{\bea}{\begin{eqnarray}}
\newcommand{\eea}{\end{eqnarray}}
\newcommand{\beas}{\begin{eqnarray*}}
\newcommand{\eeas}{\end{eqnarray*}}

% Derivatives
\newcommand{\ddt}[1]{\frac{d{ #1}}{dt}}
\newcommand{\ddtt}[1]{\frac{d^2{ #1}}{dt^2}}
\newcommand{\ddr}[1]{\frac{d{ #1}}{dr}}
\newcommand{\ddx}[1]{\frac{d{ #1}}{dx}}
\newcommand{\prt}[1]{\frac{\partial{ #1}}{\partial t}}
\newcommand{\prtc}{\left. \frac{\partial f}{\partial t} \right|_c}
\newcommand{\prx}[1]{\frac{\partial{ #1}}{\partial \vect{x}}}
\newcommand{\prv}[1]{\frac{\partial{ #1}}{\partial \vect{v}}}
\newcommand{\prxi}[1]{\frac{\partial{ #1}}{\partial \vect{x}_i}}
\newcommand{\prvi}[1]{\frac{\partial{ #1}}{\partial \vect{v}_i}}
\newcommand{\prxo}[1]{\frac{\partial{ #1}}{\partial \vect{x}_1}}
\newcommand{\prvo}[1]{\frac{\partial{ #1}}{\partial \vect{v}_1}}

\newcommand{\prxx}[1]{\frac{\partial{ #1}}{\partial x}}
\newcommand{\pryy}[1]{\frac{\partial{ #1}}{\partial y}}
\newcommand{\przz}[1]{\frac{\partial{ #1}}{\partial z}}
\newcommand{\prrr}[1]{\frac{\partial{ #1}}{\partial r}}
\newcommand{\prth}[1]{\frac{\partial{ #1}}{\partial \vartheta}}
\newcommand{\prph}[1]{\frac{\partial{ #1}}{\partial \varphi}}

% Commonly used vectors
\newcommand{\vx}{\vect{x}}
\newcommand{\vz}{\vect{z}}
\newcommand{\vk}{\vect{k}}
\newcommand{\vs}{\vect{s}}
\newcommand{\vv}{\vect{v}}
\newcommand{\vV}{\vect{V}}
\newcommand{\vb}{\vect{B}}
\newcommand{\vB}{\vect{B}}
\newcommand{\ve}{\vect{E}}
\newcommand{\vE}{\vect{E}}
\newcommand{\vj}{\vect{J}}
\newcommand{\vJ}{\vect{J}}
\newcommand{\va}{\vect{a}}
\newcommand{\vpar}{\vect{v}_\parallel}
\newcommand{\vper}{\vect{v}_\perp}
\newcommand{\eper}{\vect{E}_\perp}
\newcommand{\ampE}{\tilde{\vE}_1}
\newcommand{\ampB}{\tilde{\vB}_1}
\newcommand{\amp}[1]{\tilde{#1}}
\newcommand{\skx}{\mbox{\scriptsize{\bf k}}\cdot\mbox{\scriptsize{\bf x}}}

% Unit vectors
\newcommand{\bu}{\hat{\vect{b}}}
\newcommand{\vu}{\hat{\vect{n}}}
\newcommand{\ru}{\hat{\vect{r}}}
\newcommand{\xu}{\hat{\vect{x}}}
\newcommand{\yu}{\hat{\vect{y}}}
\newcommand{\zu}{\hat{\vect{z}}}
\newcommand{\tu}{\hat{\mbox{\boldmath $\vartheta$}}}

% Commonly used vector products
\newcommand{\vbc}{\frac{\vv\times\vect{B}}{c}}
\newcommand{\Vbc}{\frac{\vV\times\vect{B}}{c}}
\newcommand{\jbc}{\frac{\vj\times\vect{B}}{c}}

% Vector operators
\newcommand{\dv}{\nabla\dotp}
\newcommand{\rot}{\nabla\crs}
\newcommand{\Lap}{\nabla^2}

% Others
\newcommand{\qm}{\frac{q}{m}}
\newcommand{\inv}[1]{\frac{1}{ #1}}
\newcommand{\half}{\frac{1}{2}}
\newcommand{\opj}{\omega_{pj}}
\newcommand{\unten}{\mbox{\rm I}}

\title{Observations of microwave emission from solar jets and comparison with MHD simulations}

\author[0000-0003-1358-5522]{Costas E. Alissandrakis}
\affil{Deparment of Physics, University of Ioannina, Ioannina 45110, Greece}
\author[0000-0000-0000-0000]{Vasilis Archontis}
\affil{Deparment of Physics, University of Ioannina, Ioannina 45110, Greece}
\author[0000-0001-8900-5948]{Kostas Moraitis}
\affil{Deparment of Physics, University of Ioannina, Ioannina 45110, Greece}

\begin{abstract}
We computed the thermal microwave emission from a 3D magnetohydrodynamic (MHD) simulation and compared it with observations of solar jets. The simulation treats the emergence of magnetic flux into the solar atmosphere and its interaction with a low, pre-existing ambient magnetic field. This interaction leads to the formation and development of a jet, driven by an eruption. The computed 17\,GHz radio emission is compared with a number of observed jets, with respect to their morphology, their flux, and the rise time of the radio flux. We find that the MHD model reproduces the characteristics of lower-intensity jets reasonably well, whereas there are differences with stronger jets. We suggest possible ways to obtain more realistic jets from MHD simulations, so that they match better the real jets.
\end{abstract}


\section{Introduction}\label{sect:intro}
During the last few decades we have realized that the quiet Sun is not that quiet after all, with numerous small-scale energy release events occurring all over the place. These events are readily observed in spectral lines and continua in the microwave, mm, UV, EUV, and X-ray wavelength ranges \citep[e.g.][]{1973ApJ...185L..47V,1981SoPh...69...77H, 1983ApJ...272..329B, 1994ApJ...431L.155K, 1997SoPh..175..341I, 2015ApJ...813...86I, 1999A&A...341..286B, 2001SoPh..198..313D, 2014Sci...346C.315P, 2020A&A...638A..62N,2021ApJ...914...70J}. They have been given various names: coronal bright points \citep{1973ApJ...185L..47V}, brightenings, transient phenomena, explosive events \citep{1983ApJ...272..329B}, miniflares, microflares, nanoflares, UV bursts \citep{2018SSRv..214..120Y}, IRIS bombs \citep{2014Sci...346C.315P}, campfires \citep{2021A&A...656L...4B} and so on, sometimes creating confusion and making it hard to tell which is what. 

Jets \citep{1992PASJ...44L.173S} are a well-defined category of small-scale brightenings, characterized by an accompanying collimated ejection of material (the spire) from their base into the corona \citep[see the reviews by][]{2016SSRv..201....1R, 2021RSPSA.47700217S, 2022FrASS...920183S}. They are best observable in the quiet Sun and in coronal holes, but they are also numerous in active regions \citep[see, e.g., the recent work of][]{2024A&A...689L..11N}. They should not be confused with surges, which are large-scale ejections of material associated with flares \citep[see, e.g.,][]{1976sofl.book.....S}, although their origin could be due to similar mechanisms. 

Morphologically, jets are classified into {\it anemone jets}, having an inverted Y shape with a spire emerging from a bright dome-shaped base and {\it two-sided-loop jets}, which appear as a pair of plasma beams ejected in opposite directions. 
\citep{1994xspy.conf...29S}. Anemone jets are further classified as {\it standard jets}, in which the spires remain thin and narrow during the entire lifetime of the jet, and {\it blowout jets}, in which the spire broadens with time until its size becomes comparable to the width of the jet base \citep{2010ApJ...720..757M}.

There are few observations of jets in the radio range; \cite{1997ApJ...491L.121K} were the first to associate X-ray jets with 17\,GHz emissions observed with the Nobeyama Radioheliograph (NoRH), a work that was subsequently extended by \citet{1999ApJ...520..391K}. In a more recent work, \cite{2021Ge&Ae..61.1083K} compiled a catalog of jets, including NoRH data \citep[see also][]{2021STP.....7b...3A}.

The observations suggest a magnetic origin of jets, particularly one in which emerging flux reconnects with the pre-existing ambient magnetic field. 
Similarly, over the past years, a large number of 2D and 3D numerical simulations have been performed, showing that the interaction between emerging and ambient magnetic fields can reproduce some of the main properties of jets at various atmospheric heights \citep[see, e.g.][and references therein]{2016SSRv..201....1R}.
  
However, to our knowledge, the results of the numerical models have not been tested quantitatively against radio observations. In this work, we attempt such a comparison by computing the microwave emission as a function of time, based on the numerical simulation of a solar jet by \cite{2023ApJ...952...21C}. In the next section, we briefly describe the model and in Section~\ref{sect:Comput} our computations. We compare our results with observations in Section~\ref{sect:Obser} and discuss our results in Section~\ref{sect:discuss}. Our conclusions are presented in Section~\ref{sect:conclude}.

\section{The magnetohydrodynamic model}\label{sect:MHD}
The magnetohydrodynamic (MHD) model follows the evolution of an emerging magnetic flux tube, interacting with a pre-existing magnetic field. The 3D resistive MHD equations are solved using the Lare3D numerical code \citep{Arber_etal2001}. The simulation takes into account Joule and viscous dissipation, but not the radiative losses. The inclusion of partial ionization is described in detail in the work by \cite{2023ApJ...952...21C}.

{ 
The MHD model that we are using in this study includes the emergence of magnetic flux in the solar stratified atmosphere. It has been used in the past to simulate various solar events, such as eruptions, jets and flares on various scales. The model solves the MHD equations, which were normalized, expressing the density in units of 1.67$\times 10^{-4}$\,kg\,m$^{-3}$, the magnetic field in units of 300\,G, and the length in units of 180\,km. From these follow the units of all other quantities, e.g. of the velocity (2.1\,km\,s$^{-1}$), the temperature (649\,K), and the time (86.9\,s). Such a setup is typical for this kind of experiments, and simulates solar conditions to a large degree.}

The numerical box consists of a uniform 3D grid with $420^3$ points in Cartesian geometry, which corresponds to a size of $64.8\,\mathrm{Mm}$ in all directions. In the vertical $z$-direction, there is an adiabatically stratified layer that is mimicking the convective layer below the photosphere, which is located at $-7.2\,\mathrm{Mm} \leq z < 0\,\mathrm{Mm}$. The photosphere/chromosphere is located above that layer, at $0\,\mathrm{Mm}\leq z < 1.8\,\mathrm{Mm}$. Higher up we have the transition region, at heights $1.8\,\mathrm{Mm}\leq z < 3.2\,\mathrm{Mm}$. At the top, the stratified atmosphere includes an isothermal corona, with a temperature of about one million kelvin at $3.2\,\mathrm{Mm}\leq z < 57.6\,\mathrm{Mm}$. { The variation of the ambient temperature and the density with height is presented in figure 1 of \cite{2024A&A...690A.181M}.}

Initially, in the solar interior, there is a horizontal magnetic flux tube along the $y$-direction at a depth of $2.3\,\mathrm{Mm}$ below the surface. The magnetic field strength of the tube is $B_0=7882$~G, which corresponds to a value of plasma $\beta$ of 18.4. The flux tube has a uniform strong twist (but not kink unstable) and it is in total pressure and thermal equilibrium with its surrounding unmagnetized plasma. For the magnetic flux tube to rise, we impose a density deficit around its center, which makes it lighter and it starts to emerge towards the photosphere and above. During its emergence the flanks of the tube do not move and they remain at the solar interior, and thus, the emerging tube adopts an $\Omega$-loop shape. In the solar atmosphere, we have included an ambient uniform magnetic field, with a strength of $B_\mathrm{amb}=10$~G. The orientation of the field is along the positive $z$ direction (upwards) and is inclined, forming an angle of about 10 degrees with the $z$-axis at the vertical $x$-$z$ midplane. Thus, when the emerging tube and the ambient field come into contact, their relative orientation is such that the two interacting magnetic fields reconnect and create jets. Eventually, the magnetic shear along the polarity inversion line of the emerging bipolar region will form new twisted magnetic flux tubes, which will erupt and drive the so-called `blowout jets'.

The model gives the temperature, the density, the pressure, the velocity, the energy, the neutral particle fraction, the conductivity, the magnetic field, and the electric current in the 3D grid, quantities that are used in the following computations.

\section{Computation of the radio emission}\label{sect:Comput}
The radio emission of jets at short wavelengths has two components: free-free (ff) from thermal plasma, and gyrosynchrotron from accelerated particles { \citep{1965sra..book.....K,1970resp.book.....Z,2020FrASS...7...74A,2020FrASS...7...57N}}. Gyroresonant emission requires too high a magnetic field, while H$^-$ emission can be safely ignored for the wavelength range that we are considering. The MHD model does not include particle acceleration, thus our computations are limited to the free-free component. This restricts our comparison with the observations to events with a low degree of circular polarization, which is a fairly safe diagnostic of energetic particles.

For the computation of the ff emission we need the electron temperature, $T_e$ and the electron density, $N_e$; the latter is not provided directly by the MHD model, but can be computed from the gas pressure and the fraction of neutral particles.
The intensity of the emission, expressed in terms of brightness temperature, $T_b$, is
\be
T_b=\int_0^L T_e(\ell)(1-e^{-\tau(\ell)}) d\tau(\ell)  \label{eq:Tb}
\ee
where the integration is along the line of sight, $\ell$, and the optical depth, $\tau$, is given by:
\be
\tau=\frac{\xi}{f^2}\int_0^L \frac{N_e^2}{T_e^{3/2}} d\ell \label{eq:tau}
\ee
where $f$ is the frequency of observation and $\xi(f,N_e,T_e)$ is a slowly-varying factor encompassing the ionization states and the Gaunt factor, { computed from equation 6.1 of \cite{1996ASSL..204.....Z}}.

\begin{figure*}[t]
\centering
\includegraphics[width=\textwidth]{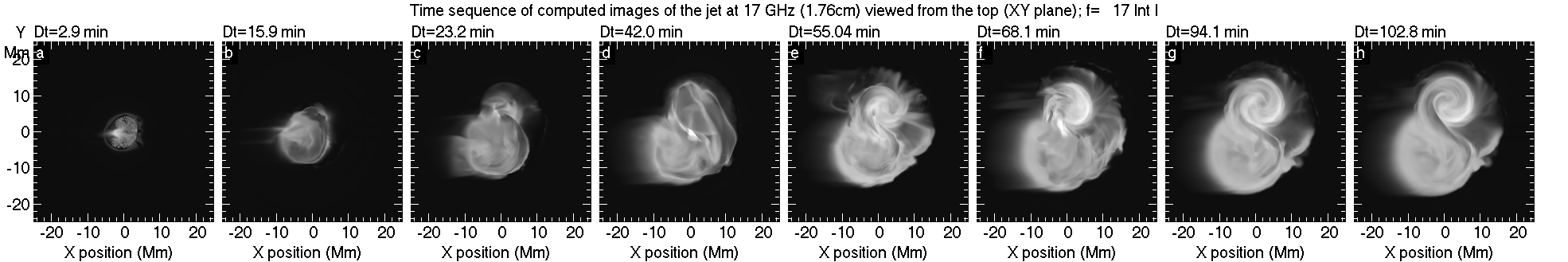}
\smallskip
\includegraphics[width=\textwidth]{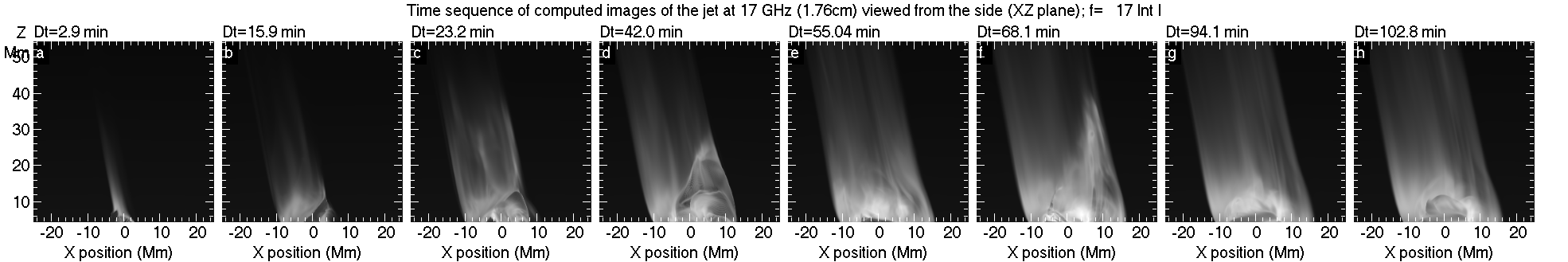}
\smallskip
\includegraphics[width=\textwidth]{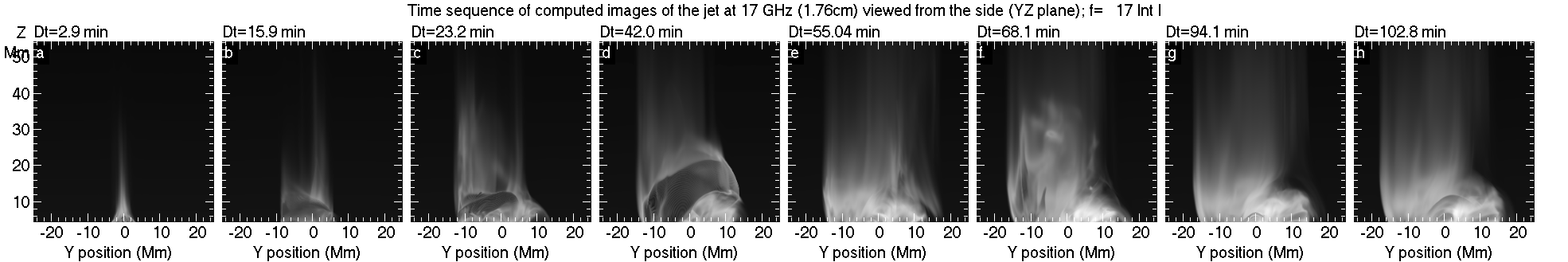}
\medskip 
\includegraphics[width=.95\textwidth]{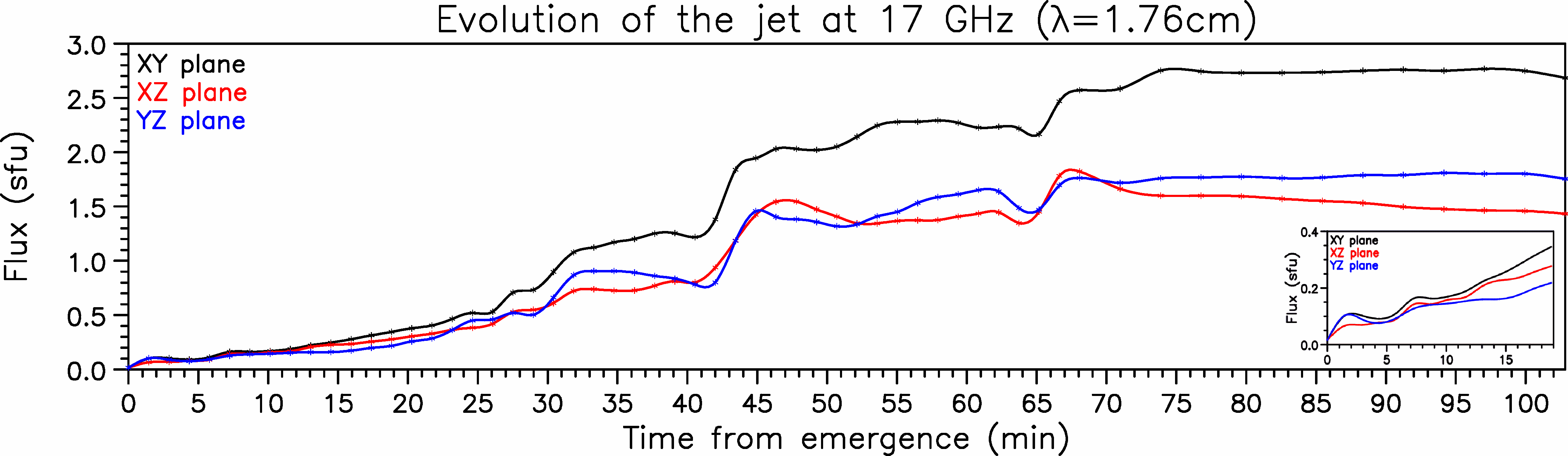}
\caption{Upper three rows: Computed images of the model jet at 17\,GHz. From top down: viewed from the top (projection on the XY plane), viewed from the side (projection on the XZ plane), viewed from the side (projection on the YZ plane). $T_b$ values range from 0 to $4\times10^5$\,K, displayed with gamma=0.35.
Bottom panel: The flux of the computed 17\,GHz emission as a function of time. The insert shows the early evolution of the jet in an expanded scale. In this and subsequent figures the full lines are spline interpolations through the actual values (asterisks).}
\label{fig:Tb images}
\label{fig:Flux}
\end{figure*}

Radio images were computed from Eq.~(\ref{eq:Tb}), by integrating along the three principal axes: $z$ (projection on the $xy$ plane, i.e. viewed from the top), $y$ (projection on the $xz$ plane, i.e. viewed from the side), and $x$ (projection on the $yz$ plane, i.e. viewed from the side). The computations were performed at 17\,GHz, for comparison with NoRH observations, and at 100\,GHz, for future comparison with ALMA Band3 data. We only computed Stokes parameter $I$ (total intensity), since Stokes $V$ (circular polarization) is very low for ff emission. As the spatial resolution of the NoRH is not adequate for the comparison of the model brightness with the observed, we also computed the flux of the emission, by integrating $T_b$ over each principal plane.

Figure~\ref{fig:Tb images} shows a time sequence of selected 17\,GHz images and the corresponding flux (bottom panel). Due to the higher opacity of the jet plasma as seen from above (XY plane), the flux is higher than for the side views. The 100\,GHz images and flux are similar to the respective 17\,GHz ones.

We first note that the images in Figure~\ref{fig:Tb images} describe well the evolution of the jet, as predicted by the MHD model. After the emergence of the flux tube into the corona, reconnection gives rise to a standard jet (panel a in Figure~\ref{fig:Tb images}). Eventually, blowout jets are emitted from the emerging flux region. Panel b in  Figure~\ref{fig:Tb images}, shows the first blowout jet, while panels (c, d, e) and (f, g, h) show the evolution of the next two blowout jets. Notice that the blowout jets have a much wider spire compared to the standard jet and that their base (see the size of the emission at the top row) is larger.

We have estimated that the lifetime of the jets is around $6.5-7.5$~min. For simplicity we define the lifetime of the jets as the time during which high velocity upflows in the corona increase the background temperature of the plasma. The whole process, from the formation of the erupting flux rope until the initial emission of the blowout jet and it's decay, takes much longer.  

In Figure~\ref{fig:Flux} we note that, $\sim70$~min after the emergence, the flux stabilizes. This happens because no other jet occurs in the simulation. The last blowout jets starts at around $68$~min.

\section{Observations for comparison with the MHD model}\label{sect:Obser}
Practically all the available information on jet microwave emission has been obtained with the NoRH \citep{1994IEEEP..82..705N}, which has a nominal spatial resolution of 10\arcsec\ at 17\,GHz, although the actual resolution is usually worse than that. In this work we used information from \citet{1997ApJ...491L.121K} and \citet{1999ApJ...520..391K}, but the bulk of our jets were selected from the catalog of \citet{2021Ge&Ae..61.1083K}, which covers the period from May 17, 2010 to May 5, 2022. 

\begin{figure}%[b]
\centering
\includegraphics[width=.47\textwidth]{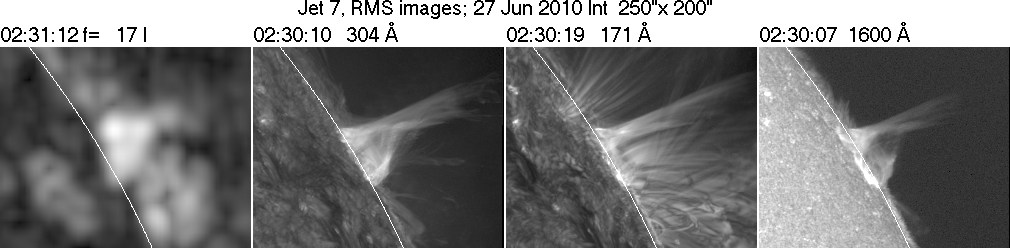}
\bigskip\includegraphics[width=.47\textwidth]{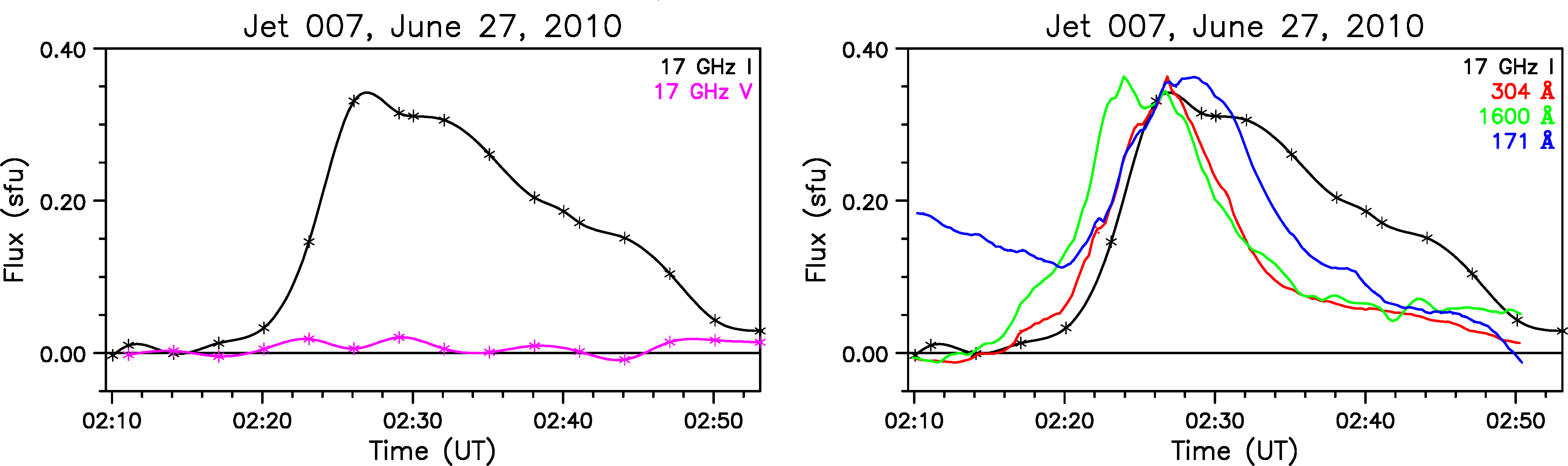}
\caption{Top row: Images of the RMS of the intensity variation for jet K007. From left to right: 17\,GHz; AIA 304\,\AA; AIA 171\,\AA; AIA 1600\,\AA. The white arc marks the photospheric limb. Bottom row: The flux as a function of time at 17\,GHz, Stokes I and V (left), and at all wavelength bands (right). In the plot at right this and in subsequent similar plots the flux of AIA bands is arbitrary units, with its peak normalized to the 17\,GHz peak.}
\label{fig:007}
\end{figure}

We selected events near the limb, where jets are easy to detect, as they project against the sky; an additional selection criterion was that the events should be well visible in full-disk images available at the NoRH site. For events before 2015 we obtained 3-min cadence 17\,GHz images from the NoRH site\footnote{https://solar.nro.nao.ac.jp/norh/images/3min/} and, for later events, 10-min cadence images\footnote{https://solar.nro.nao.ac.jp/norh/images/10min/}. SDO/AIA images in the 1600, 304, and 171\,\AA\ bands, formed at chromospheric, low transition region, and low corona temperatures, respectively, as well as STEREO/SECCHI-EUVI images in the 195\,\AA\ band were used in conjunction.

Given the low resolution of the NoRH images, it is not meaningful to compare the observed brightness temperature with the model. We therefore restricted our comparison to the flux, its evolution with time, and the morphology of the jets.
Our results are presented in this section.

\subsection{Jet K 007}
This jet was observed above the NW limb on June 27, 2010. STEREO A images show that it was a two-sided loop jet, and that part of its base was behind the limb as seen from the Earth. Instead of showing its evolution with a sequence of images, we give in Figure~\ref{fig:007} images of the root mean square (RMS) of the intensity, computed at every pixel of the field of view over the duration of the event; this type of display provides a concise picture and facilitates the comparison among frequency bands. 

We note the similarity of radio emission of the jet base with that at 304\,\AA\ and 1600\,\AA, although the spire is less prominent at 17\,GHz, apparently due to the inferior resolution. In the 171\,\AA\ RMS image the base is less prominent; moreover, this image is contaminated with features associated with time varying coronal loops.

Light curves of the jet are shown in the lower part of Figure~\ref{fig:007}. We note that the 1600\,\AA\ emission is the first to rise and reach maximum, while the radio emission peaks at about the same time as the 304\,\AA\ emission and a second 1600\,\AA\ peak, and the 171\,\AA\ emission peaks about 2\,min later. Moreover, the 17\,GHz emission lasts longer, with a slower decay. We further note that the circular polarization is too low for any other mechanism than ff emission.

\subsection{Jet K 094}
\begin{figure}%[h]
\centering
\includegraphics[width=.47\textwidth]{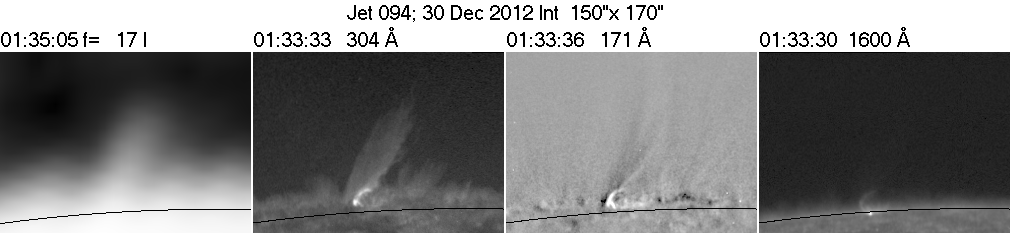}
\bigskip\includegraphics[width=.47\textwidth]{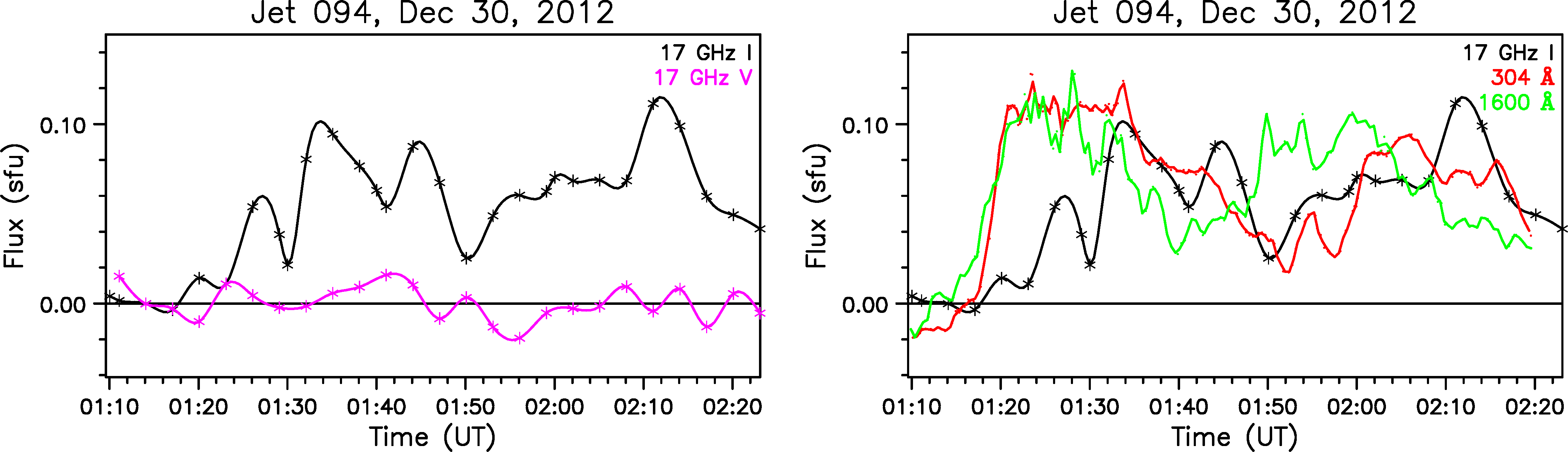}
\caption{Top row: Selected images of jet K094. From left to right: 17\,GHz; AIA 304\,\AA; AIA 171\,\AA\ with the pre-event intensity subtracted; AIA 1600\,\AA. The black arc marks the photospheric limb.
Bottom row: Left: the flux as a function of time at 17\,GHz, Stokes I and V. Right: same with 304\,\AA\ and 1600\,\AA\ added.}
\label{fig:094}
\end{figure}

K 094 is a polar jet, which occurred at a latitude of $\sim85$\degr\ (Figure~\ref{fig:094}); it is a complex event with multiple ejections of material, the full description of which is outside the scope of this work. The bright point in the 1600\,\AA\ image shows that its base was not in the back side of the disk, but the 304\,\AA\ image indicates absorption of its radiation by the foreground chromospheric plasma; the same is probably true for 17\,GHz.  

The jet is very weak in the NoRH data, with a peak flux of just about 0.1\,sfu above the background of 704\,sfu (bottom of Figure~\ref{fig:094}). Nevertheless, its form is similar to that at 304\,\AA, while at 171\,\AA\ most of the spire shows up in absorption.

\subsection{Jet K 097}
\begin{figure}[h]
\centering
\includegraphics[width=.47\textwidth]{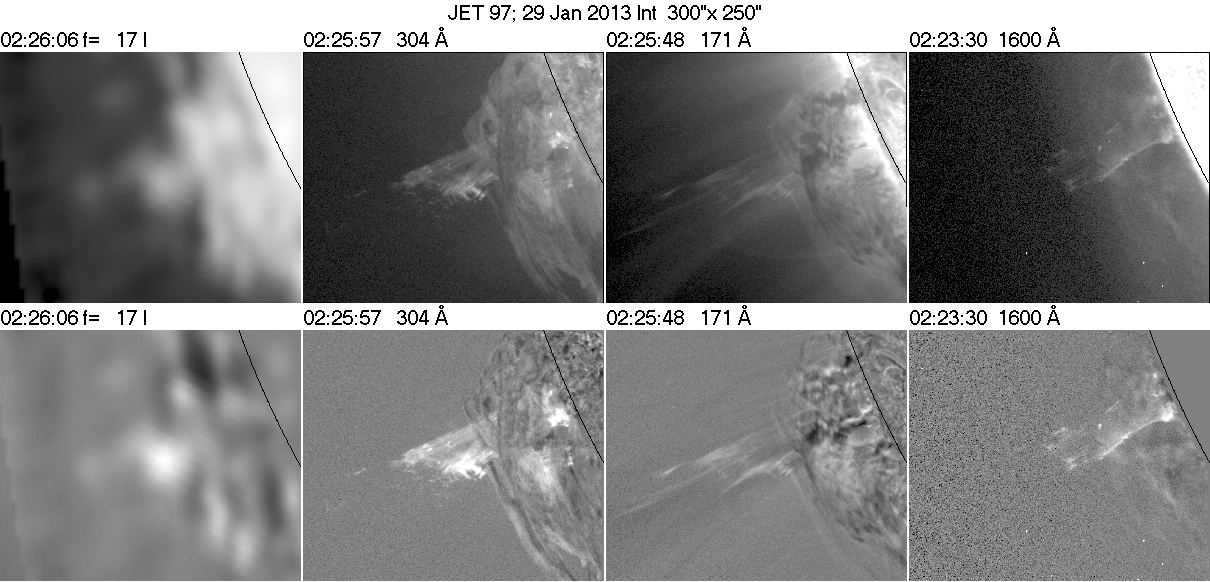}
\bigskip\includegraphics[width=.47\textwidth]{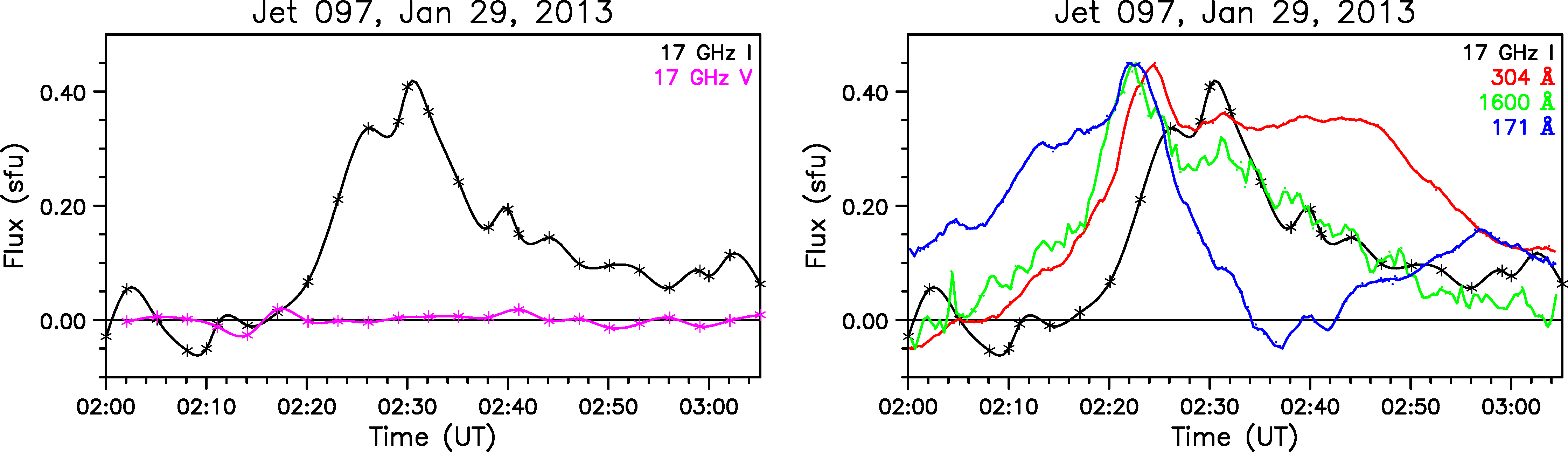}
\caption{Top row: Selected images of jet K097. From left to right: 17\,GHz; AIA 304\,\AA; AIA 171\,\AA; AIA 1600\,\AA. Middle row: the same images, after subtraction of the emission before the event. The black arc marks the photospheric limb. Bottom row: The flux as a function of time at 17\,GHz (left) and at all wavelength bands (right).}
\label{fig:097}
\end{figure}

Selected images of jet 97 of the Kaltman catalog are presented in Figure~\ref{fig:097}. STEREO B images reveal a two-sided loop jet, with its base behind the limb as seen from the Earth; moreover, a prominence at the limb hides a good part of the spire, with the effect being less severe in the 1600\,\AA\ band, where the opacity is lower. 

Light curves are shown in the bottom row of Figure~\ref{fig:097}. Due to the obscuration effects, the measured peak flux is a lower limit of the true value, while some time variations could originate in the prominence. We further note that there was no detectable circular polarization in the radio emission. 

\begin{figure}%[b]
\centering
\includegraphics[width=.47\textwidth]{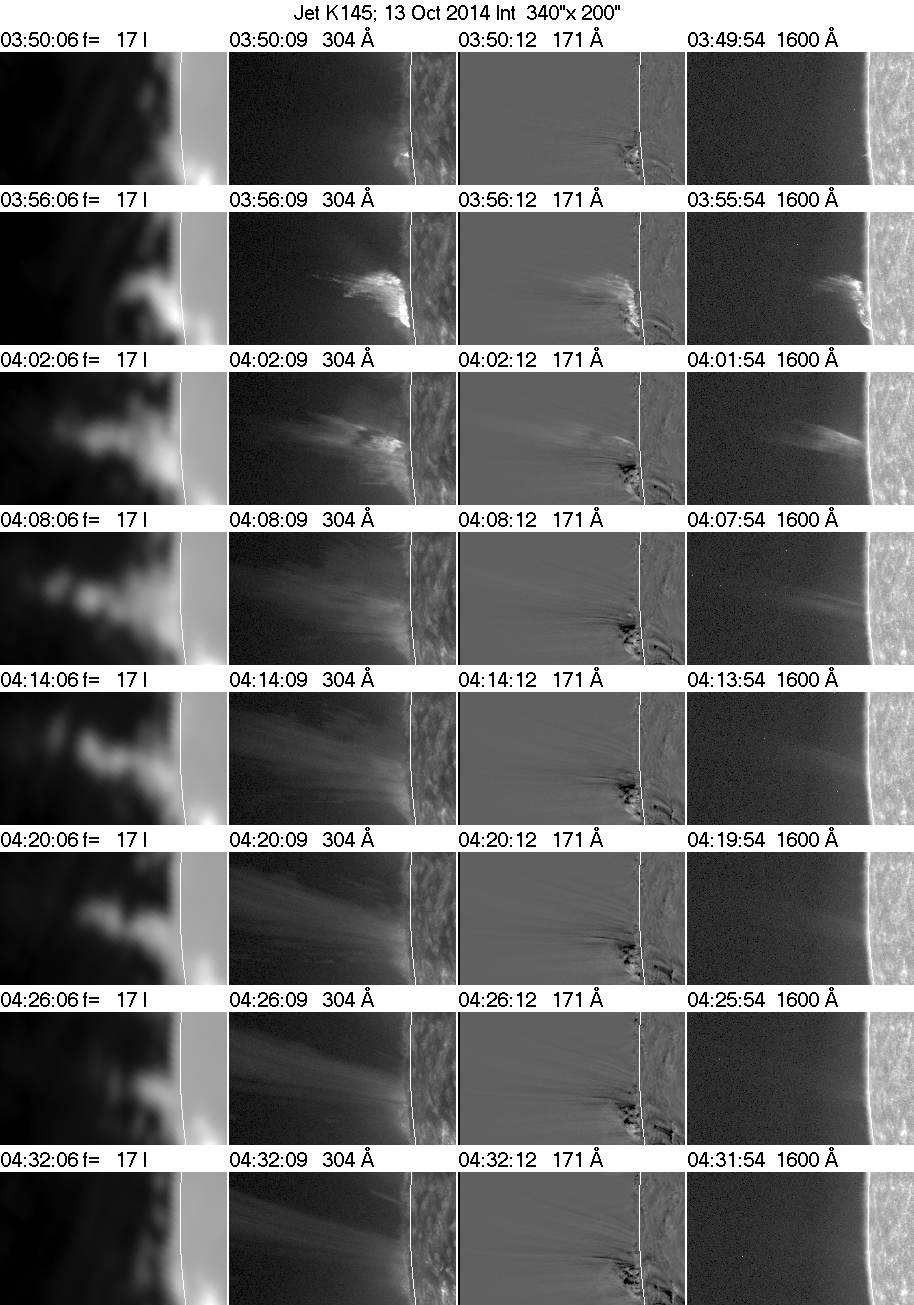}

\medskip\includegraphics[width=.47\textwidth]{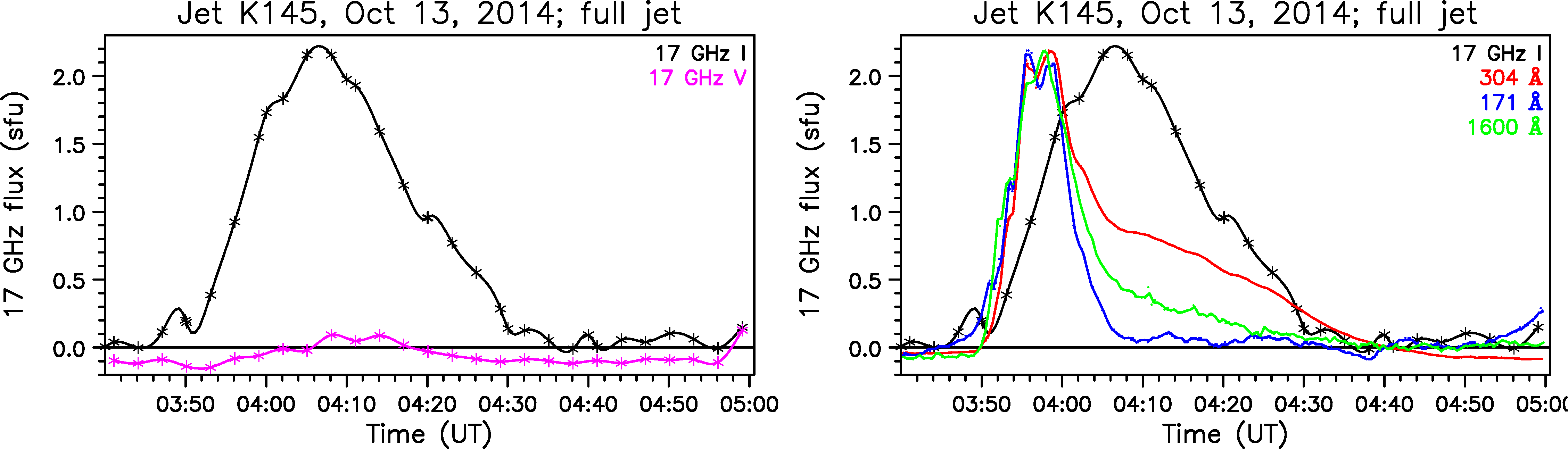}
\caption{Upper rows: Images of jet K145. From left to right: 17\,GHz, intensity; AIA 304\,\AA; AIA 171\,\AA\ with the preflare image subtracted;  AIA 1600\,\AA. The white arc marks the photospheric limb. Last row: Time profiles of the jet at 17\,GHz (left) and at all wavelength bands (right).}
\label{fig:145}
\label{fig:TimProf_145}
\end{figure}

\subsection{Jet K 145}
Figure~\ref{fig:145} shows the evolution of jet 145 from the Kaltman catalog at 17\,GHz and in the 304\,\AA, 171\,\AA\ and 1600\,\AA\ AIA bands; the color table is the same in all images of each column. The NoRH data did not show any detectable circular polarization, pointing again towards ff emission. This jet was not associated with an active region, and a part of its base was behind the limb. 

The NoRH images appear similar to the 304\,\AA\ images, taking into account the much better spatial resolution of the latter. In the 171\,\AA\ band, where the preflare intensity has been subtracted, the jet is less prominent; absorption is seen after its initial development, apparently due to cooler material in front of the emitting plasma. At 1600\,\AA\ the jet is weaker than at 304\,\AA, with a similar morphology.

The flux of the jet as a function of time at 17\,GHz is plotted in the left panel of the bottom row of Figure~\ref{fig:TimProf_145}; the right panel of the same figure shows the flux in the AIA bands, superimposed on the 17\,GHz flux and normalized to the same value range for comparison. We note that EUV emission peaks practically at the same time in all AIA bands, reflecting primarily the time evolution of the base of the jet. At 17\,GHz, where the emission from the top of the jet is comparable to that of its base, the maximum occurred $\sim10$~min after the AIA maximum. The decay of the 17\,GHz emission is notably slower than initial decay of the EUV emission, but the overall duration of the jet at 17\,GHz ($\simeq40$~min), is comparable to that at 304\,\AA; the duration is shorter in the 171\,\AA\ light curve, probably due to the presence of absorbing material.

\subsection{Jet K 199}
\begin{figure}%[h]
\centering
\includegraphics[width=.47\textwidth]{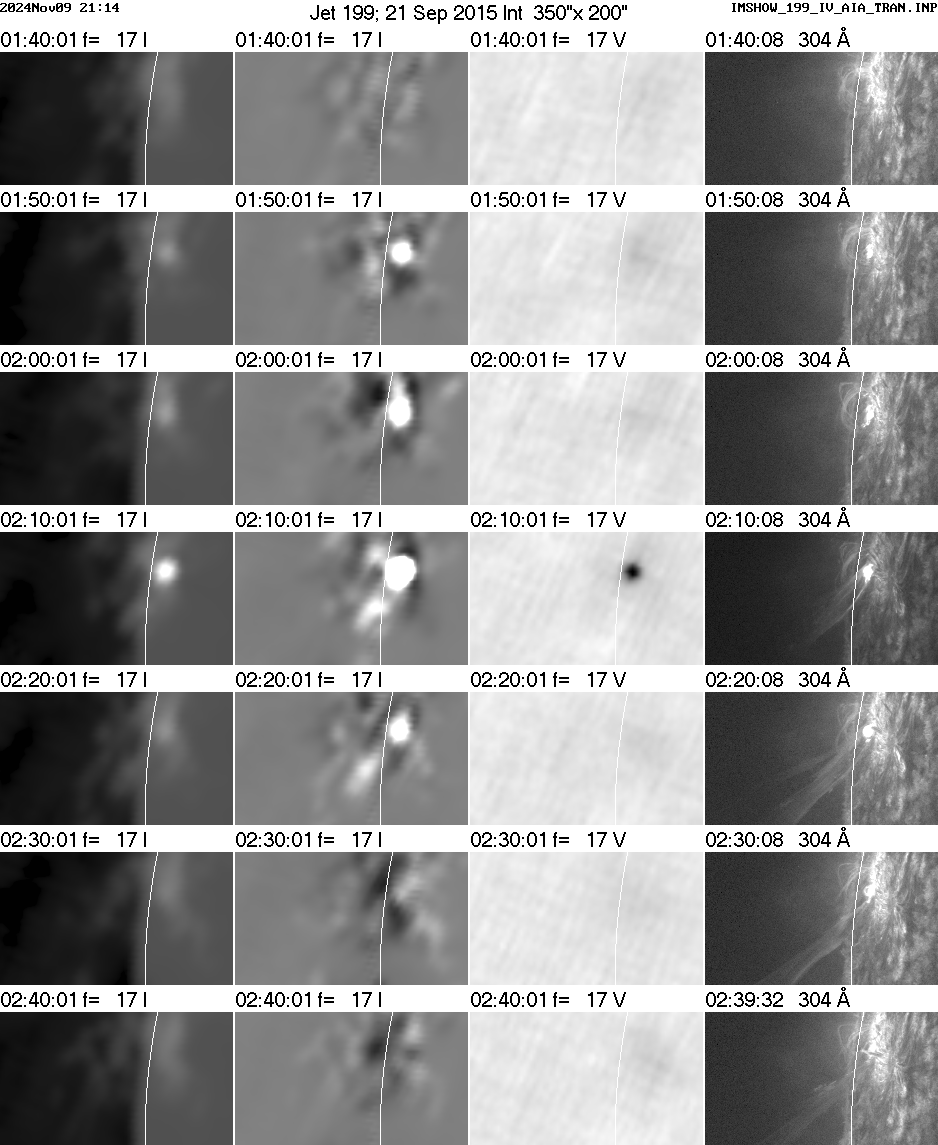}

\medskip\includegraphics[width=.47\textwidth]{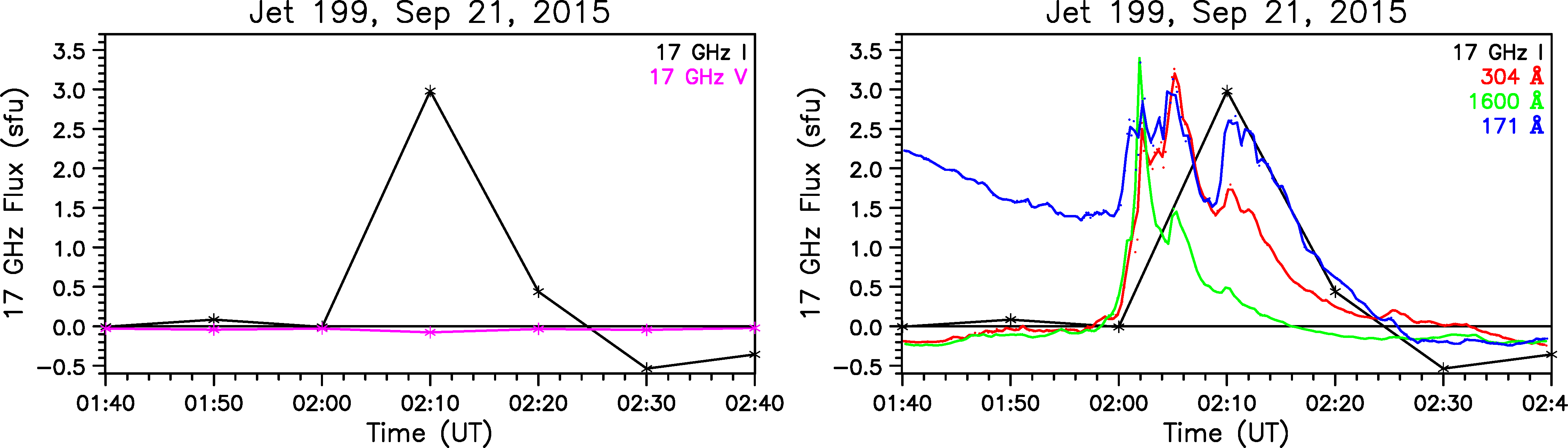}
\caption{Upper rows: Images of jet K199. From left to right: 17\,GHz, intensity; 17\,GHz intensity with pre-flare image subtracted; 17\,GHz, circular polarization; AIA 304\,\AA. The white arc marks the photospheric limb. Last row: Time profiles of the jet at 17\,GHz (left) and at all wavelength bands (right).}
\label{fig:199}
\label{fig:TimProf_199}
\end{figure}

A time sequence of images of jet 199 from the Kaltman catalog, which occurred near the east limb in AR 12420, is presented in Figure~\ref{fig:199}. In addition to the 17\,GHz images (original and after subtraction of the preflare emission), we give images in circular polarization (Stokes $V$) as well as AIA 304\,\AA\ images. Although the $V$ image shows a discrete source at the base of the jet, the degree of circular polarization is only 1\%, still in favor of pure ff emission. The value of $V$ provides an estimate of $\sim30$\,G for the magnetic field.

The light curve of the jet at 17\,GHz is shown in the left panel of the bottom row of Figure~\ref{fig:TimProf_199}; it is shown again in the right panel, together with the curves for the AIA bands at 304\,\AA, 1600\,\AA\ and 171\,\AA. The cadence of the AIA images (24\,s for 1600\,\AA, 12\,s for the others) allows the detection of several peaks, something which is not possible with the low cadence of the NoRH images (10\,min). The duration of the event in radio and EUV is comparable, but it is difficult to establish a time shift between the two.

\begin{figure}%[b]
\centering
\includegraphics[width=.47\textwidth]{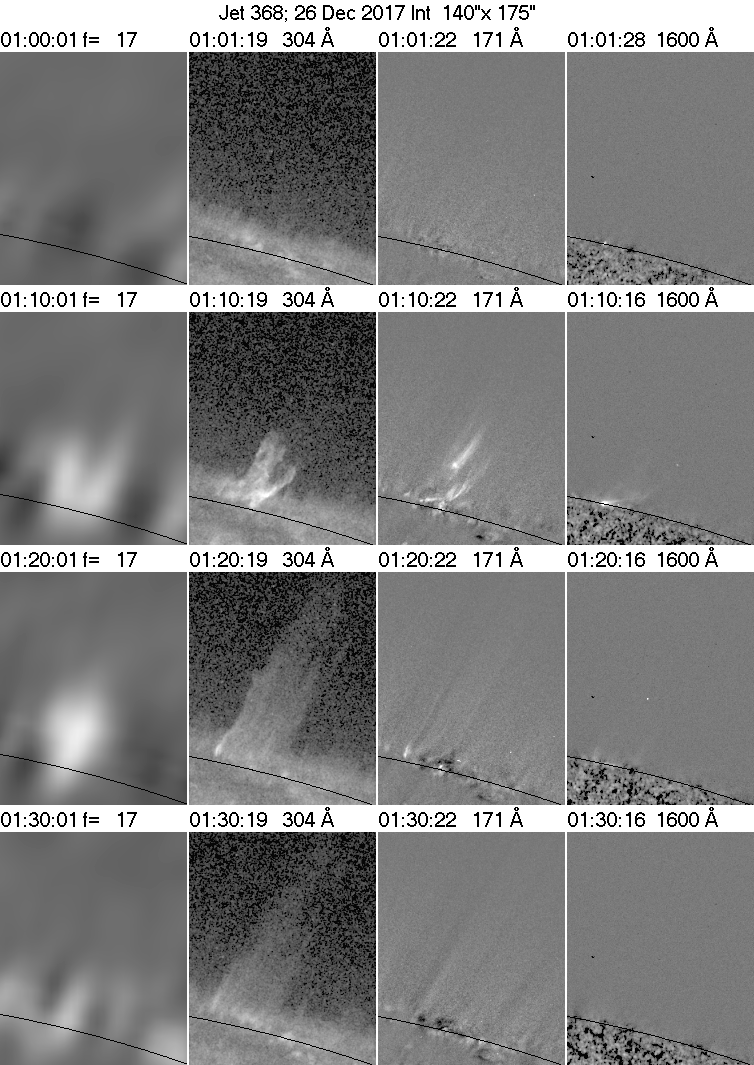}

\medskip\includegraphics[width=.47\textwidth]{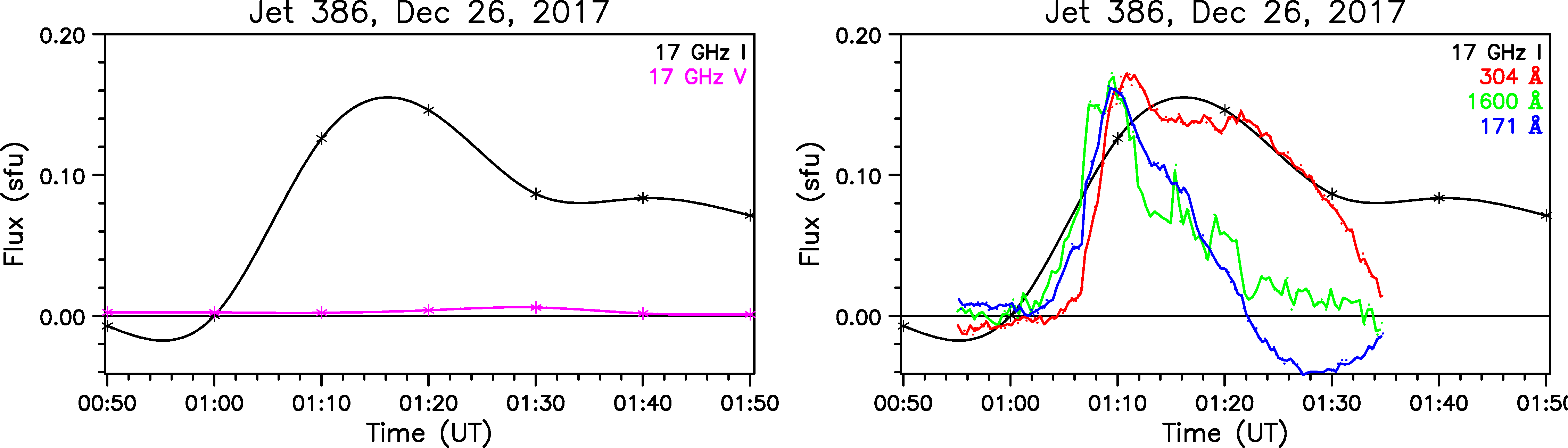}
\caption{Upper rows: Images of jet K368. From left to right: 17\,GHz intensity; AIA 304\,\AA; AIA 171\,\AA\ AIA 1600\,\AA. Images before the event have been subtracted, except for 304\,\AA. The black arc marks the photospheric limb. Last row: Time profiles of the jet at 17\,GHz (left) and at all wavelength bands (right).
}
\label{fig:368}
\label{fig:TimProf_368}
\end{figure}
\medskip
\subsection{Jet K 368}
Images of jet K368 are shown in Figure~\ref{fig:368}. The 1600\,\AA\ images show that the projected position of the base of the jet was very close to the photospheric limb, but still on the visible side of the Sun; we could, however, have some absorption from the foreground chromospheric plasma. The jet was not on the visible side of the Sun as seen from STEREO A.

Time profiles of the jet are given in the bottom row of Figure~\ref{fig:TimProf_368}. Emission was first detected in the 1600\,\AA\ band, followed by 171\,\AA\ and 304\,\AA; the latter may have been affected by chromospheric absorption. We further note that the flux curve at 171\,\AA\ reveals absorption of the background coronal emission by the jet material during the decay phase. Due to the low cadence of the NoRH images, we cannot comment on the time shift beween the 17\,GHz and the EUV emissions.

\bigskip
\bigskip

\section{Discussion}\label{sect:}\label{sect:discuss}
\begin{figure*}[ht]
\centering
\includegraphics[width=.85\textwidth]{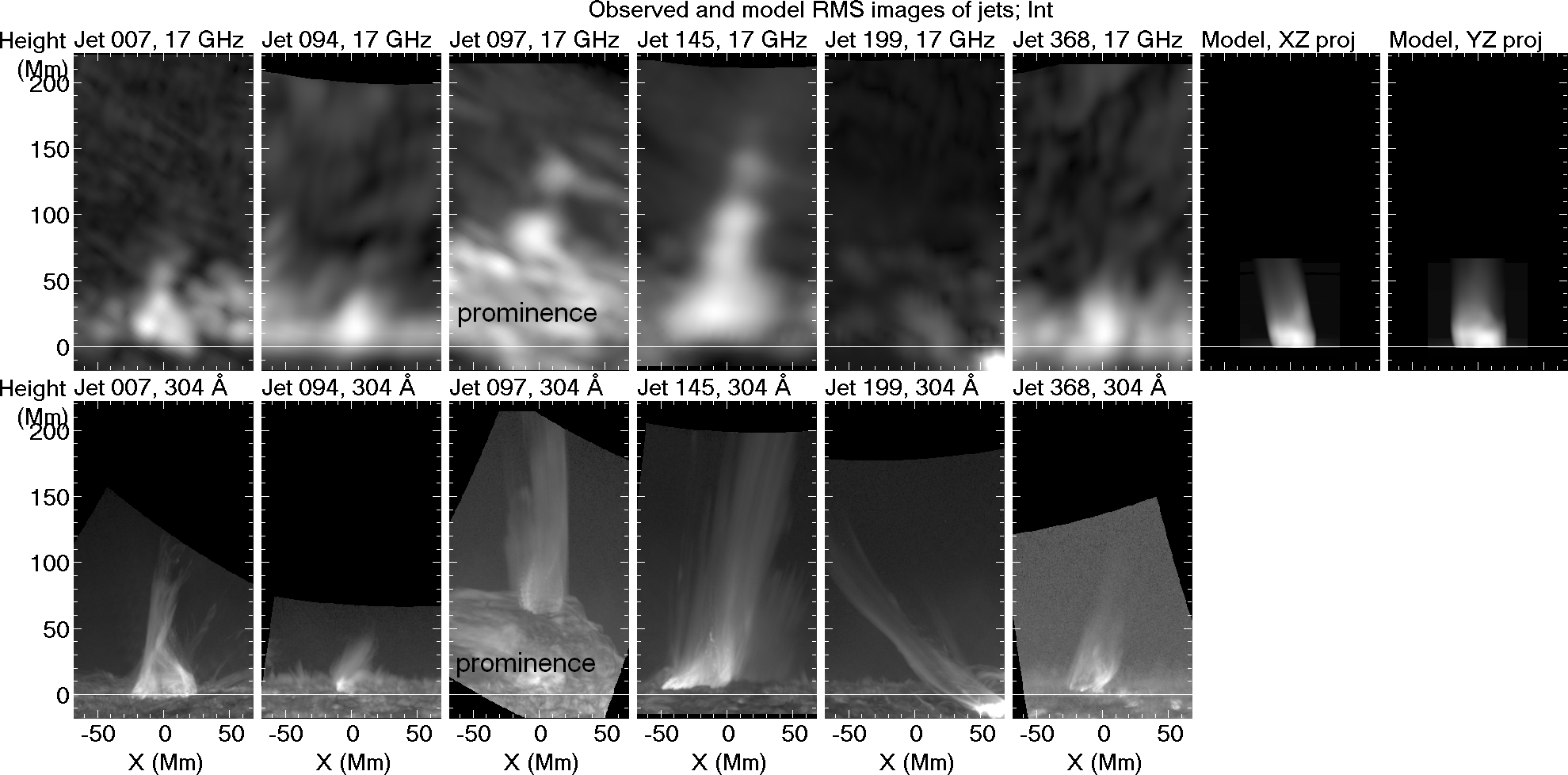}
\caption{Top row: Images of the RMS of the intensity variation for all observed jets and for side views of the MHD model (last two columns) at 17\,GHz. Bottom row: RMS images at 304\,\AA. Coordinates are in Mm and heights are measured from the photospheric limb.
}
\label{fig:RMS_POL}
\end{figure*}

We start our discussion with the comparison of the morphology of observed and model jets. The images in the top row of Figure~\ref{fig:RMS_POL} show all our jets, in the form of RMS images; RMS images of the MHD model intensity, smoothed to the NoRH resolution are given at the right. For comparison, RMS images in the 304\,\AA\ band are shown in the bottom row of the figure.

The real jets come in a variety of forms, reflecting differences in physical conditions. At 17\,GHz their height ranges from about 45\,Mm for K094 to about 170\,Mm for K145; they extend higher in the 304\,\AA\ images (bottom row of Figure~\ref{fig:RMS_POL}), apparently an effect of the better spatial resolution, better dynamic range and higher opacity. On the other hand, the model height is limited to the $\sim60$~Mm vertical extent of the computation box. We further note that, compared to the base, the spire is less bright and more wide in the model than in the observations. Finally, the model predicts a width smaller or comparable to the observed.

For a quantitative comparison, we summarize our 17\,GHz measurements and those of previous authors in Table~\ref{table:jets}. The tabulated heights are visual estimates and the rise times are time differences between the peak and the start of the jet; their accuracy is of the order of the image cadence, 3\,min for jets 007 to 145 and 10\,min for the others. In the first three rows we give results from \cite{1997ApJ...491L.121K} and \cite{1999ApJ...520..391K}. The last three rows give the predictions of the MHD model at thee phases, 10, 20, and 45\,min after the emergence of the magnetic flux tube; the tabulated flux refers to the side views.

We note that the range of peak flux values of \cite{1997ApJ...491L.121K} and \cite{1999ApJ...520..391K} are at the low side of our values. We further note that 10\,min after the magnetic flux emergence, a time interval comparable to the observed rise time, the MHD model predicts a 17\,GHz flux of 0.15\,sfu. This value is consistent with the observed one for the weak jets; however, the flux of the stronger jets in our sample is about a factor of 15 above that value, and even a factor of 2 greater than the model prediction 45\,min after emergence.

In one case (K145) we were able to measure the apparent velocity of rise at 17\,GHz. Our value (195\,km\,s$^{-1}$) is higher than that of the single case reported by \cite{1999ApJ...520..391K}, but practically identical to the model prediction for the early stages of the jet evolution; we note that the model predicts a lower ascending velocity later on.

\begin{table*}[ht]
\caption{Parameters of observed and model jets at 17\,GHz.}\label{table:jets}
\begin{tabular}{lcccccccl}
\hline
ID   &   Date     &   UT  &Location&Flux&Rise time&Height&Velocity  & Comments\\
     &            &       &        & sfu& min     & Mm&km\,s$^{-1}$\\
\hline
Ku 1997&Mar 31, 1995&03:54&Disk&0.004 - 0.011&2 - 8\\
       &Aug 25, 1992&04:06& Limb&0.08 - 0.010&$\sim4$\\
Ku 1999&1992 - 1995&&&0.08 - 0.35&&&~\,55& 19 events\\
       &           &&&           &&&(1 case)\\
\hline
K007 &Jun 27, 2010& 02:30 & N29W89 & 0.32 & ~\,9 & ~\,65\\
K094 &Dec 30, 2012& 01:40 & N86E36 & 0.10 & 10 & ~\,45\\
K097 &Jan 29, 2013& 02:30 & S25E88 & $>0.42$& ~\,9 & 150 &&Base hidden\\
K145 &Oct 13, 2014& 04:00 & S03E89 & 2.30 & 16& 170&195\\
K199 &Sep 21, 2015& 02:10 & N03E90 & 3.00 & 10& ~\,70\\
K368 &Dec 26, 2017& 01:20 & N76W80 & 0.15 & 10& ~\,60\\
\hline
\multicolumn{4}{l}{Model, 10 min after flux emergence}&0.15&&50&195\\
\multicolumn{4}{l}{Model, 20 min after flux emergence}&0.28&&20\\
\multicolumn{4}{l}{Model, 45 min after flux emergence}&1.41&& $>70$ &110\\
\hline
\end{tabular}
\end{table*}

\section{Summary and conclusions}\label{sect:conclude}

In this work we have compared the observed radio emission of selected jets at 17\,GHz with the 
{ 3D MHD flux emergence simulation of \cite{2023ApJ...952...21C}. We chose the specific experiment as a generic example of flux emergence simulations and examined its relevance to observed jets. }
Due to the low resolution of the radio images, we limited our comparison to the morphology, the flux and the rise time; as the model does not fully account for energy dissipation, we could not compare the duration of the jets.

We found that the model reproduces reasonably well the characteristics of low intensity jets, with 17\,GHz flux $\lessapprox0.2$\,sfu. Stronger jets have higher flux and extend higher than the model predicts.

Considering the broad range of physical parameters in real jets, we believe that our results are encouraging, as far as the validity of the MHD model is concerned. By changing the initial parameters of the MHD model, it should be possible to obtain a better agreement with a larger set of observed cases.

As an example, a higher jet height could be obtained by extending the computational volume with a corresponding increase in computational burden. We could also get higher jet densities by changing the normalisation quantities (magnetic field, density and length) of Lare3D. We conclude that studies similar to our work are necessary in the future for MHD models to become more realistic and closer to the observed jets.

\begin{acknowledgments}
This research has been supported by the European Research
Council through the Synergy grant No. 810218 (`Whole
Sun', ERC-2018-SyG). The authors express their gratitude to colleagues and staff of NoRH, STEREO/SECCHI-EUVI and AIA/SDO for making their data publicly available.
\end{acknowledgments} 

\bibliographystyle{aasjournal}
\bibliography{References,More}

\end{document}